\documentstyle[11pt]{article}
\input epsf
\begin{document}
\centerline{\large \bf Future Measurements at RHIC}
\vspace{0.5cm}

\centerline{Richard Seto}
\centerline{ University of California, Riverside}
\centerline{ Riverside, CA 92521, USA}
\vspace{0.5cm}
\centerline{\it Proceedings of the Nuclear Dynamics Workshop, Nassau, Bahamas, January 2002}
\begin{abstract}
I review the present status of measurements at RHIC and suggest
possible upgrades which will be necessary to answer some of the most critical
questions in Relativistic Heavy Ion Physics.
\end{abstract}
\section{Introduction}

The Relativistic Heavy Ion Collider was completed in 1999 and a host
of new experimental results soon appeared. During the past year, the
nuclear physics community has undertaken the task of formulating a
long range plan to guide funding priorities for the next 10 years.
What follows is my view of the future. I have chosen to be rather
speculative at times. I will first review what we have learned so far,
then pose the questions formulated in the Long Range Plan (LRP), and
use these to guide a discussion of the future measurements which
should be made.

In addition to the many recent experimental results, theory has been
making rapid progress in explaining the phenomena we are seeing at
RHIC.  This is possible for two reasons. First, powerful theoretical
techniques, have been brought to bear on the problem. These are
reviewed elsewhere in these proceedings by A Dimitru\cite{dimitru}.
Secondly, RHIC has taken us into a regime in which these theoretical
techniques are applicable.  It is the convergence of theory and
experiment which leads to new understanding.  The collection of
experimental data alone makes us simple accountants of miscellaneous
facts. On the other-hand a non-verifiable theory may be elegant, but
will not give us an understanding of the world in which we live. I
would like to suggest that a sort of standard model is just beginning
to emerge in our understanding of the evolution of relativistic heavy
ion collisions in the regime we have entered.  This picture is not
entirely new. The general picture of the phase diagram of QCD has been
around for many years.  What is new, is that RHIC, now takes us into a
regime in which theoretical techniques can now be used to do
reasonably reliable calculations, using the model, of experimentally
measurable quantities.  As such we are beginning to enter into an era
of precision measurements in which theoretical predictions of a
fundamental nature can be made, and experiments can be done under
controlled conditions so that these ideas can be tested.

To do many of these measurements, RHIC will need an increase in
luminosity.  What is being proposed by the community is an increase by
a factor of 40 by the use of electron cooling.  In addition,
experiments will need to be upgraded, to enhance capabilities
and increase sensitivity for rare or hard to measure probes,
and to increase the ability of detectors to handle a higher data rate
through the use of higher level triggers and expanded DAQ.
The details of this will be
covered in the next talk by T. Ludlam\cite{ludlam}.

\section{What Have We Learned?}\label{learned}

I will first summarize of some of the results from RHIC as they
stand now. Much of this material is covered in more depth in separate
contributions to this volume. Unless specified, the data is
from the first run where the center-of-mass energy was 130 GeV. A
small amount of data is now available from run-2 where the center of
mass energy was at the canonical value of 200 GeV. Needless to say,
the various RHIC collaboration are working hard to produce the results
from this latest run.

\subsection{Initial Conditions: Energy Density and the Colored Glass 
Condensate}\label{init}

The first task is to make an experimental estimate of the initial
energy density of the system formed in collisions at RHIC to see
whether it is high enough to induce a phase transition.  Lattice
calculations indicate that the phase transition should occur between
150 and 200 MeV corresponding to energy densities between 0.6 and 1.8
GeV/fm$^3$ with the lower estimate being favored by models which
include light fermions.  Bjorken has given an estimate of the energy
density assuming boost invariance\cite{bjorken}
$$\epsilon=\frac{E}{Volume}=\frac{1}{\pi R^2} \frac{1}{2c\tau_0}
({2\frac{dE_T}{dy}})$$
where R is the radius of the nucleus in
question, and $\tau_0$ is the time after the initial collision of
thermalization when a temperature can first be defined, typically
taken to be between 0.2 and 1 fm/c.  As theory has improved it has
been recognized that there are at least two relevant time scales for
$\tau_0$.  The first, as described below is a formation time related
to the Colored Glass Condensate. The second, is the actual
thermalization time.  The PHENIX collaboration has measured $dE_T/dy
\sim dE_T/d\eta = 503 \pm 2\ GeV$, giving an energy density times
thermalization time to be $\epsilon\cdot\tau_0 =4.6\ 
(GeV/fm^3)(fm/c)$. Conservative estimates give $\tau_0 \sim 1 fm$,
whereas estimates from the equilibration of gluons give $\tau_0 \sim
0.2 -0.3 fm$.  This gives an energy density estimate of between 5 and
23 GeV/fm$^3$, an order of magnitude greater than that required for
the phase transition as predicted on the lattice.

Recently, McLerran\cite{CGC} and his colleagues have proposed that the
initial conditions in heavy ion conditions could be calculated
assuming that the gluon distributions at low x are saturated. The
gluon distributions at low-x for protons $\sim 1/x^\delta$, hence
violate unitarity at very low x and saturate.  In protons it will
occur at about $x\sim 10^{-4}$ depending on the Q$^2$.  In a Lorentz
contracted nucleus, there is essentially a geometric magnification
factor of A$^{1/3}$ making the relevant $x\sim 10^{-2}$.  McLerran and
his colleagues assumed that since the occupation numbers were high,
one could use a classical approximation. This leads to
non-perturbative calculations of various experimentally measured
quantities, using renormalization group methods which depend on a
scale Q$_s$, the saturation momentum given by $Q_s=\alpha_s
\frac{xG_A(x,Q_s^2)}{\pi R_A^2}$. At RHIC, this is around 1-2 GeV/c.
They have dubbed this state, a Colored-Glass-Condensate for reasons I
will not explain here.

\begin{figure}[htb]
  \begin{center}
   \mbox{\epsfxsize 6in \epsfbox{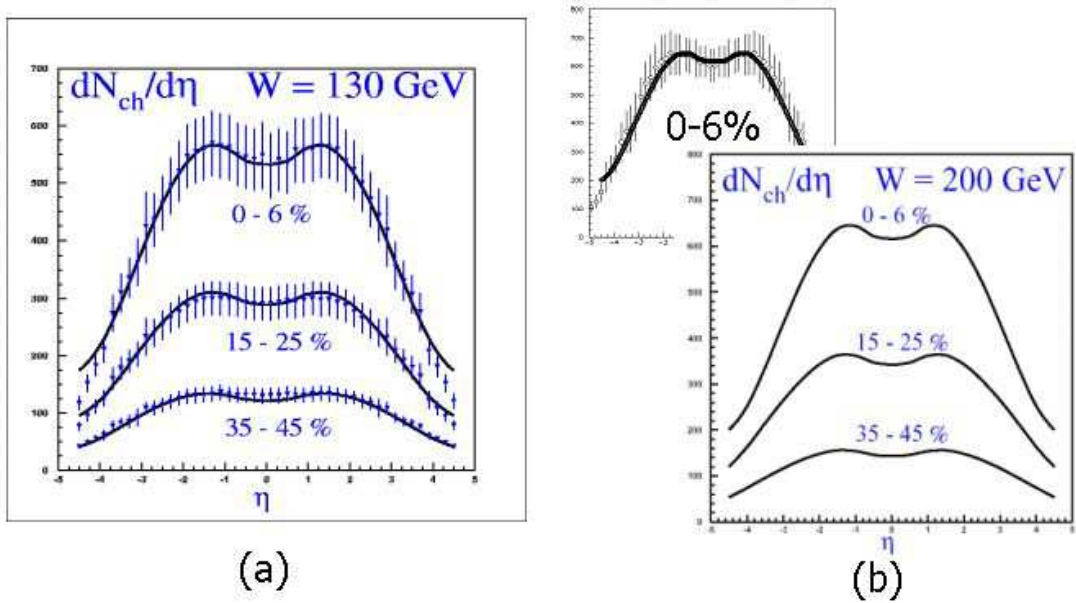}}
  \end{center}
  \caption{(a) dN/d$\eta$ distributions as measured by the PHOBOS 
collaboration and compared with the model of Kharzeev and Nardi
for $\sqrt{s}$=130 GeV  
(b) Theoretical prediction by Kharzeev and Levin
 for $\sqrt{s}$=200 GeV, where the inset 
shows the most-central bin as compared with the PHOBOS measurement.}
  \label{fig1}
\end{figure}

One of the first calculations to be done using these methods was for
the multiplicity distributions as a function of
centrality\cite{kharzeevnardi} at $\sqrt{s}=$130 GeV (figure
\ref{fig1}a).  Admittedly this was a post-diction after the data was
available. However, Kharzeev and Levin\cite{kharzeevlevin} then
predicted the centrality dependence at the higher energy of
$\sqrt{s}=$200 GeV.  The prediction, shown in the right panel of the
figure does extremely well for the recently published central data
from the PHOBOS collaboration.  It will be important to observe
whether the centrality dependence at the higher energy is borne out.
The model gives us an initial energy density of 18 GeV/fm$^3$ in good
agreement with the estimate above. This corresponds to a formation
time of $\tau_0 \sim \tau_{formation} \sim 1/Q_S \sim 0.1-0.2 fm/c$.
Presumably equilibration happens somewhat later. This time corresponds
to the time in which the gluons ``go on mass shell'', hence this
energy density refers not to a thermalized energy density, but rather
the maximum energy density reached by the system. Kharzeev, Levin, and
Nardi\cite{kln} have also made a prediction at $\sqrt{s}=$22 GeV, but
they point out that the model may not work since $\alpha_s$ becomes
large. Their central prediction at this energy is that the number of
charged particles per participant at mid-rapidity rises slowly by
about 20\% for central collisions due to the running of the coupling
constant $\alpha_s$.
 
\begin{figure}[htb]
  \begin{center}
   \mbox{\epsfxsize 3in \epsfbox{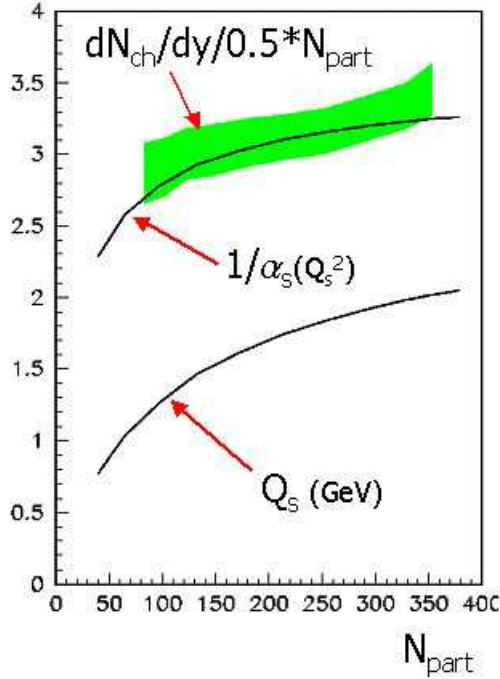}}
  \end{center}
\caption{A plot illustrating the centrality dependence of the CGC
  model showing Q$_s$ vs N$_{part}$, together with the corresponding
  values of 1/$\alpha_s(Q^2_s)$ and the PHOBOS multiplicity data.  
The important issue here are the
  dependencies of the various quantities on the centrality, and not
  their actual values.}
  \label{fig11}
\end{figure}

It is worth examining the source of the centrality dependence of the
multiplicity in the CGC model. In this model the number of gluons per
participant is proportional to the inverse of the strong coupling
constant hence, somewhat schematically, $N_{ch} \sim N_{gluons} \sim
1/\alpha_s(Q^2_s)$. The value of $Q^2_s$ rises as a function of
centrality since the quantity of interest is the number of gluons per
unit area in the target nucleus as seen by a particular projectile.
The more the number of gluons, the higher the saturation scale.  This
is shown in figure \ref{fig11} where $(Q_s)$ is plotted versus
N$_{part}$. The corresponding value of $1/\alpha_s(Q^2_s)$ is also
plotted and compared to the multiplicity per participant as measured
by the PHOBOS collaboration. 
The shapes of the distributions are similar since $dN/d\eta$
calculated in the CGC model is proportional to $1/\alpha_s(Q^2_s)$.
This gives rise to the agreement between the theory and the data shown
in figure \ref{fig1}a.  This result connects the centrality dependence
of the the multiplicity per participant to the running of the coupling
constant. If $\alpha_s$ did not run, there would be no centrality
dependence.  It is interesting that a bulk property such as the total
multiplicity is intimately connected with a fundamental property of
QCD - the running of the coupling constant.  The calculation relies on
the fact that the number of produced particles is essentially equal to
the number of gluons liberated in the initial collision, which in turn
is equal (actually a factor of 2ln2=1.34\cite{kov}) to the number of
gluons in the initial virtual saturated state.  This then implies that
the number of particles may be conserved through the parton to hadron
transition - a miraculous fact noted in reference
\cite{kharzeevnardi}. One can speculate why this is so. An
understanding of this may lead to insights into the chiral and
deconfinement phase transitions, since the process of partons turning
into hadrons includes the generation of the dressed masses
of the quarks which comprise hadrons and the confinement of quarks
into hadrons.  Such a conservation may not hold for other quantities
such as the momentum distributions of the final state particles which
may be influenced by later stages of the system.

The Colored Glass Condensate model nicely explains one of the
seemingly disappointing early results from RHIC.  The charged particle
multiplicity was found to be lower than naive expectations
which sprung from entropy arguments invoking the fact there would be
more degrees of freedom in a system of quarks and gluons than in a
system of hadrons. In the CGC scenario, the production of entropy is
simply choked off by the saturation.

\begin{figure}[htb]
  \begin{center}
   \mbox{\epsfxsize 6in \epsfbox{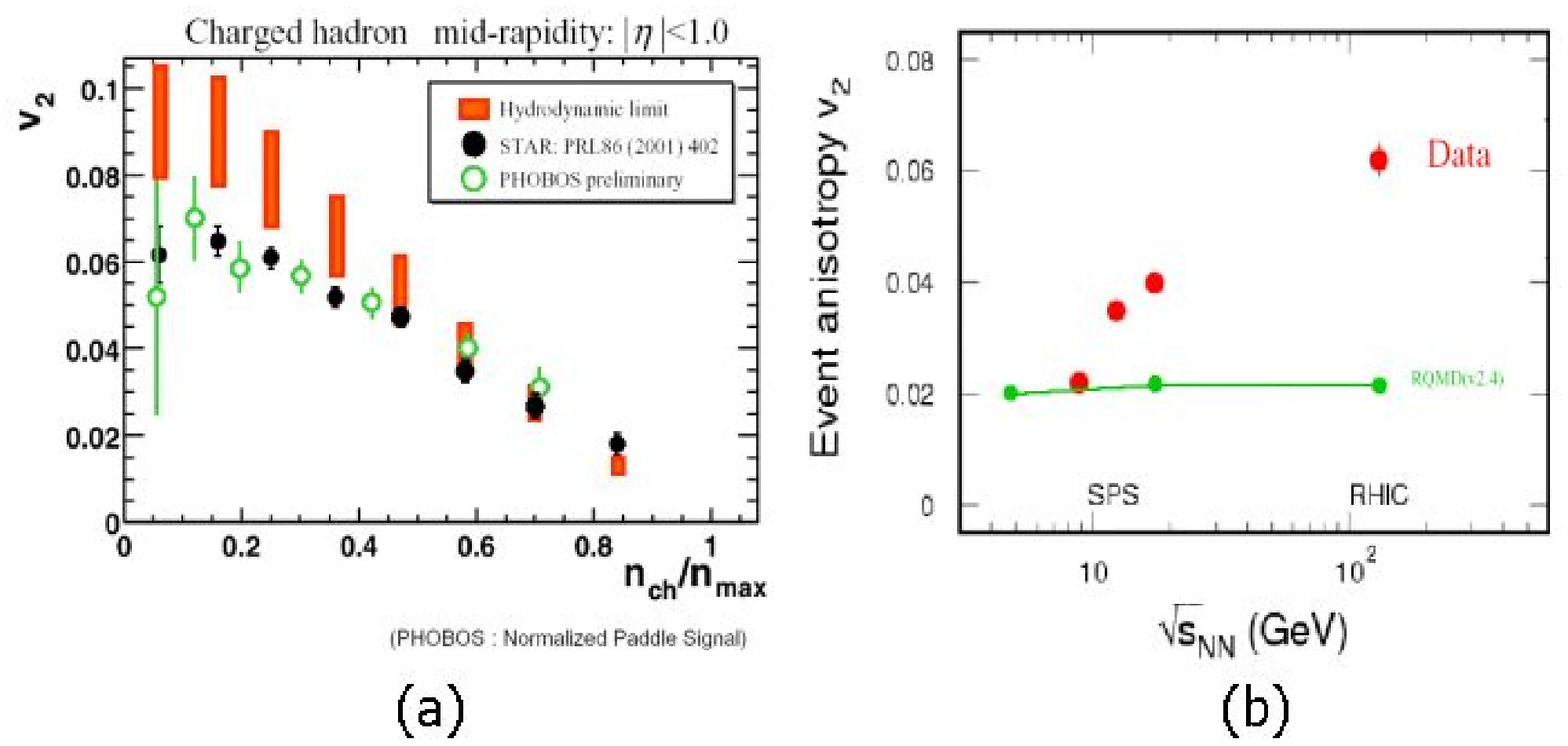}}
  \end{center}
\caption{(a) The event anisotropy v$_2$ as a function of centrality 
measured by the total number of charged particles; $n_{ch}/n_{max} \sim 1$ 
is central. v$_2$ is a measure of the strength of the elliptic flow. 
(b) v$_2$ as measured as a function of $\sqrt{s}$,showing the large increase 
from SPS to RHIC.}
  \label{fig2}
\end{figure}

Early results from both the STAR and PHOBOS collaborations indicated
that there was a surprisingly strong elliptic flow in non-central
collisions (see figure \ref{fig2}).  Elliptic flow is an anisotropy
which exhibits itself in the momentum distribution of the produced
particles. This anisotropy in momentum reflects a space anisotropy of
the original, almond shaped collision region.  For the space
anisotropy to be transfered to the momentum, the system must
equilibrated. Since flow is a collective effect, and each particle's
direction is defined by the spacial distribution of all the particles,
each particle must ``know'' the shape of the original distribution.
The effect is self quenching.  If the time
for equilibration is too long, the collision region looses its almond
shape, hence the momentum distribution looses its anisotropy.  A strong
elliptic flow, therefore, implies an early thermalization. The details
of this argument were developed in a more rigorous fashion by U. Heintz 
and his colleagues\cite{sorge}. One can see in figure \ref{fig2}a, that in
mid-central collisions, the data gives a value that is consistent with
hydrodynamics which of course assumes thermalization with a zero mean
free path. This hydrodynamical model\cite{kolb} requires a
thermalization time of 0.6 fm/c giving an energy density of 20
GeV/fm$^3$, again consistent with the previous estimate.

\begin{figure}[htb]
  \begin{center}
   \mbox{\epsfxsize 6in \epsfbox{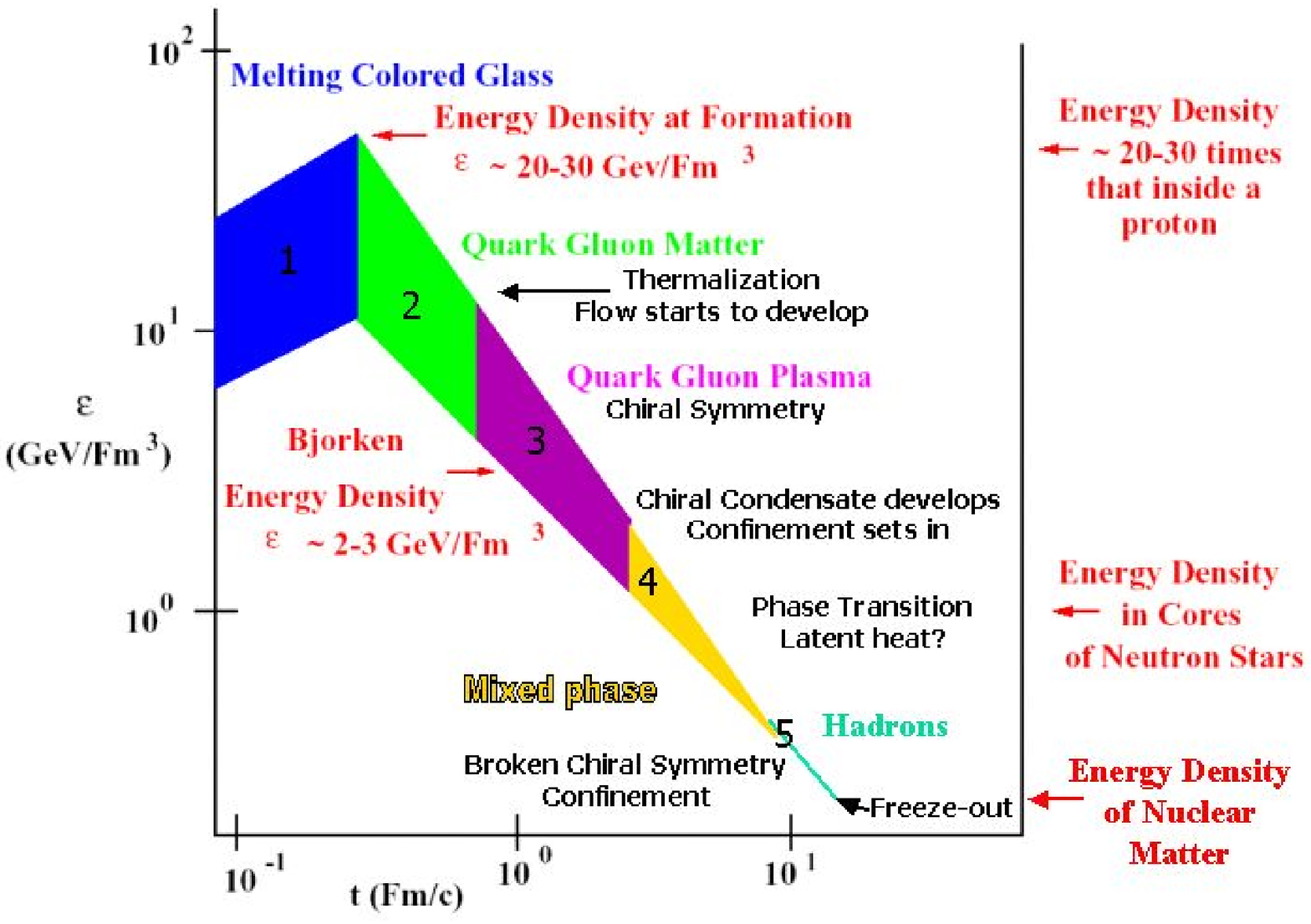}}
  \end{center}
\caption{A schematic time history of the energy density in RHIC Au-Au 
collisions. In the first stage the nuclei collide and the low-x virtual
gluons become ``real''. Thermalization occurs in the second stage and leads
to the third stage which is presumably the thermalized Quark-Gluon Plasma. 
During this stage the system undergoes
3-dimensional expansion and flow can start to develop. This fourth
stage is the mixed phase where hadronization occurs. Particles freeze-out
soon thereafter at about $\tau \sim 10$fm/c.}
  \label{fig3}
\end{figure}

It seems fairly certain, then that the initial state formed in Au-Au
collisions at RHIC is a strongly interacting system at very high
energy density. We can begin to assemble a model (figure \ref{fig3}) 
of RHIC collisions
beginning as a 
Colored Glass Condensate - essentially a saturated condensate of
gluons. This is a virtual state since it
is presumably there in all nuclei. What is important in collisions at
RHIC, is that the colliding system provides a mechanism for the
virtual gluons to become ``real'' in a formation time which is of
order 0.1-0.2 fm/c.  These ``real'' gluons, then form the initial
state which is of interest to us.  At this point they are not
equilibrated - they have just come into existence at the beginning of
the second stage.

The mechanism and timescale of thermalization is probably the most
critical question to answer in the short term. If thermalization of
gluons occurs then we have the inevitable conclusion that we have
created a Quark-Gluon Plasma. The Parton Cascade Model\cite{PCM}
showed the importance of radiative processes in equilibration.
Recently, Mueller and his colleagues\cite{mueller} have shown this by
analytical techniques as well. They give an estimate for the
equilibration time as 1/Q$_s$ multiplied by some power of $\alpha_s$.
The Monte-Carlo implementation of the Parton Cascade Model gives an
equilibration time of $\sim$ 0.3 fm/c, for a total formation
time plus thermalization time of about 0.5 -0.6 fm/c.  This is
consistent with the value found from the elliptic flow results and
hydrodynamic models of 0.6 fm/c, however this agreement should
probably not be taken too seriously until more reliable estimates for
the thermalization time can be made.

We are left then, at a time of about 0.6 fm/c with a plasma of
primarily gluons. Presumably quarks begin to appear as the chemical
reactions continue. One of the interesting characteristics of this
state is that value of the quark condensate $<q\overline{q}>$ is zero
and chiral symmetry is restored. The question of how the
condensate actually develops is an interesting one which we know very
little about.  The equilibrated system, which we can now characterize
as a QGP expands and generates flow as it cools. The system proceeds
through a mixed phase as it begins to hadronize. This is arguably
the most interesting time in the history of the collision - the 
third and fourth stages in figure \ref{fig3}.

At this point we are faced with host of questions.  Is the transition
first or second order, or simply a cross over?  We do not know the
relationship between the deconfinement transition, responsible for the
binding of quarks into hadrons, and the chiral transition, responsible
for the generation of hadronic mass. Are the transitions at the same
value of the temperature?  What is the latent heat? Is there an
inflation related to the phase transition as is presumably true in the
early universe? Are there large scale fluctuations?  Answering these
questions are the primary goals for the present and future program at
RHIC.

\subsection{Hadrons-the Later Stages of the Collision}\label{eqil}

The system continues to cool and expand in the hadronic phase until
eventually all the interactions cease and the particles freeze-out. It
is these hadrons which we can actually detect. They should reflect the
state of the system at hadronic freeze-out.  RHIC has made impressive
progress in the measurements of these hadronic observables. More
complete descriptions are elsewhere in these
proceedings. I will only give a brief overview.

\begin{figure}[htb]
  \begin{center}
   \mbox{\epsfxsize 6in \epsfbox{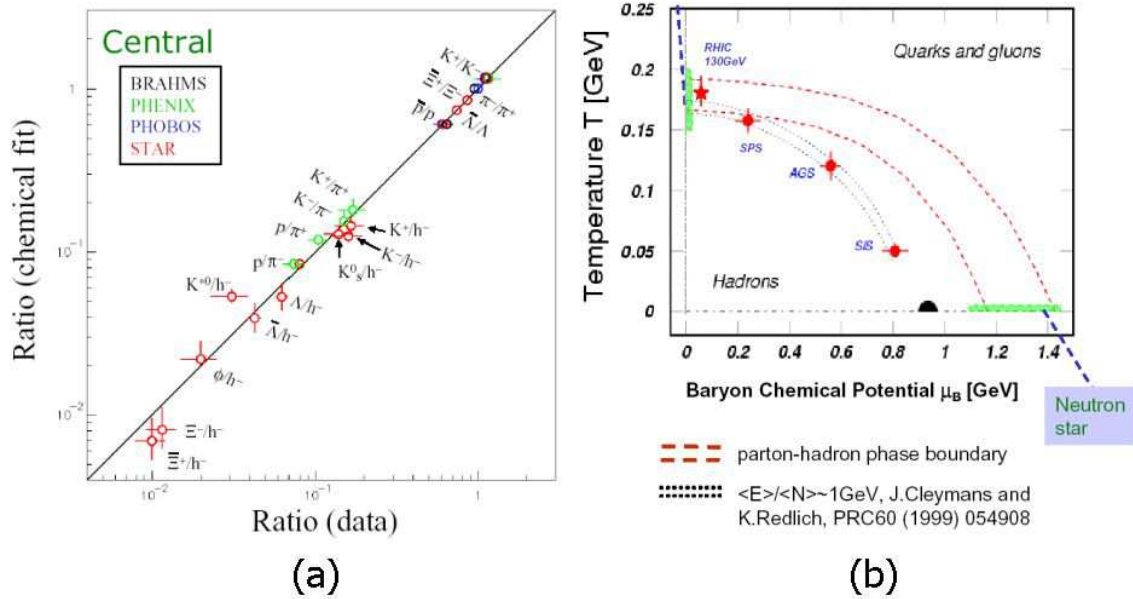}}
  \end{center}
\caption{(a) Ratios of various particle yields measured at RHIC as compared
  to the values in the model assuming chemical equilibration, using a
  single freeze-out temperature T, the baryo-chemical potential $\mu_B$,
  and a strangeness saturation parameter $\gamma$. The best fit is to
  $\gamma_s \sim 1$.  (b) Freeze-Out values of systems measured at various
  collision energies on a phase diagram.}
  \label{fig4}
\end{figure}

Figure \ref{fig4} shows the plethora of particle ratios that have
already been measured. These values are compared to a model assuming
chemical equilibrium with three parameters: the freeze-out temperature
T, the baryo-chemical potential $\mu_B$, and a strangeness saturation
parameter, $\gamma_s$. First, one observes that the model works reasonably well
and unlike at the SPS, strangeness is saturated, i.e. $\gamma_s=1$.  The
fit gives a temperature of 190 MeV and a baryo-chemical potential of
0.04 at chemical freeze-out. This is consistent with a near baryon free
system - a fact which can be deduced from the value of
$\overline{p}$/p of 0.6 measured by several of the RHIC
collaborations.
 
\begin{figure}[htb]
  \begin{center}
   \mbox{\epsfxsize 6in \epsfbox{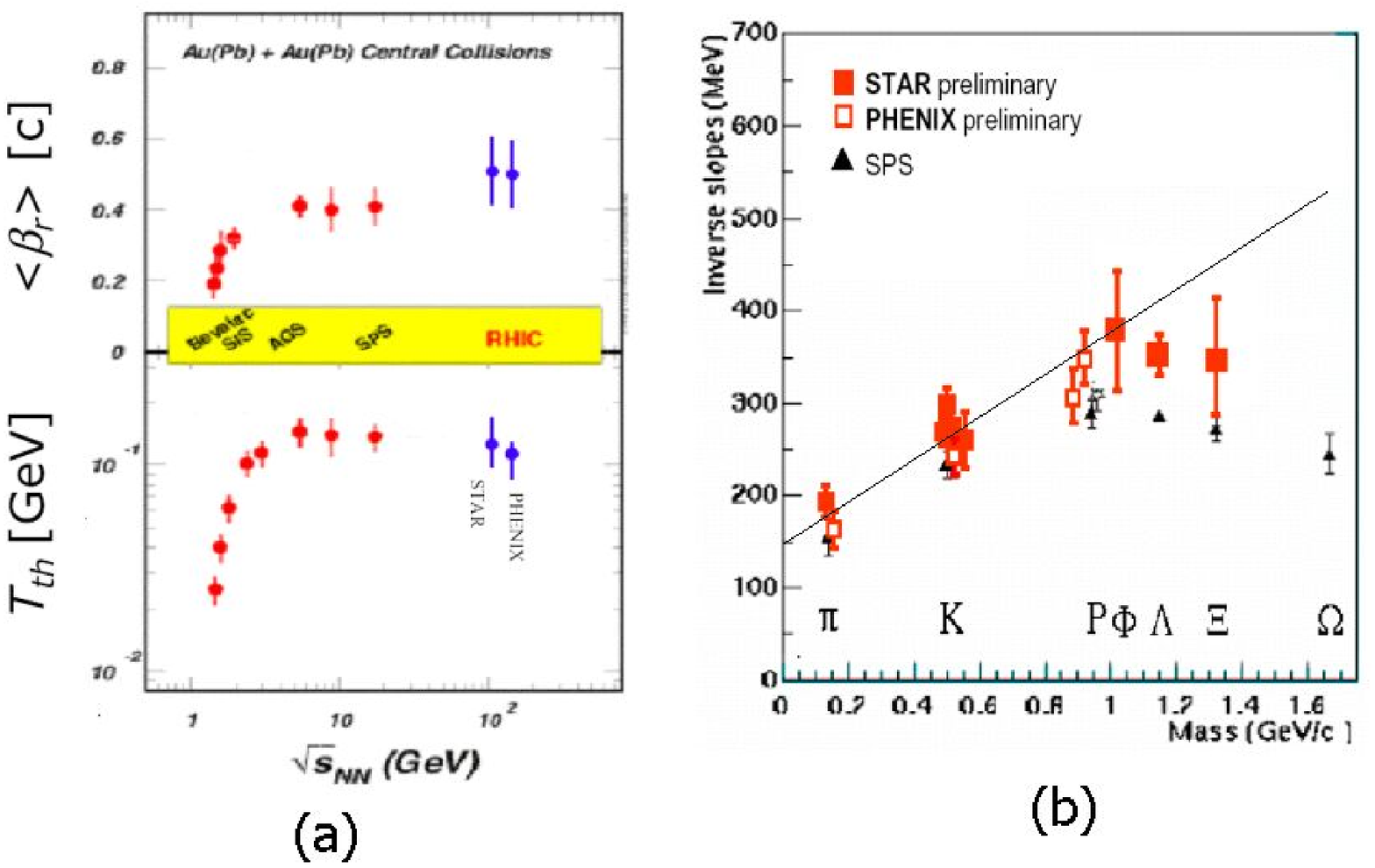}}
  \end{center}
\caption{(a) Parameters for a fit to inverse slopes of a 
variety of particles, in a model assuming kinetic equilibrium, a single 
kinetic freeze-out temperature T$_{th}$, and a radial flow velocity, 
$\beta_r$. 
(b) The inverse slopes of a variety of particles. }
  \label{fig5}
\end{figure}

We can look at thermal equilibrium as well. Figure \ref{fig5}a
shows a fit to the inverse slopes, for central collisions in a model
assuming a single freeze-out temperature and radial flow for a variety
of center of mass energies from GSI to RHIC.  One observes that the
temperature appears to saturate at a value of about 130 MeV at the AGS
and remains the same at the SPS and RHIC. At RHIC however, the value
of $\beta_r\sim 0.5$ is noticeably larger. (I note that the error bars for
$\beta_r$ and T$_{th}$ are correlated.  If one held T$_{th}$ constant the
errors on the flow velocities would be considerably smaller. This can
be seen by simply examining the original inverse slopes.) It appears
that at RHIC we are observing an increasing radial expansion.  A
intriguing speculation is that this expansion might be the result of
``inflation'' caused by the chiral phase transition, similar to that
which caused the inflationary phase of the early universe.

One of the questions that one might ask is whether this rapid radial
expansion develops in the early gluon phase of the system, during a
mixed phase, or later during the hadronic phase before freeze-out.
Figure \ref{fig5}b, shows the actual inverse slopes measured for a
variety of particles as a function of the mass. The strange-baryons
show a tendency to be below the line corresponding to a particular
temperature and radial flow velocity. This might indicate that these
particles were freezing out somewhat earlier then their lighter
counterparts at a time when the flow had not had time to develop
fully. A similar trend was seen at the SPS.  This interpretation would
imply that at least some part of the radial flow developed during a
hadronic stage.
 
\subsection{Penetrating Probes}\label{probes}

Let us return to the early stages of the collision. Are there probes
which can provide information about the state of the system and
substantiate the ideas outlined earlier, that is to study the state
which may be the QGP? Of course the answer is yes - these are the
penetrating probes - leptons, photons, and most notably high p$_T$
particles-presumably the leading particles of jets.

Perhaps the most striking result from the recent data at RHIC comes
from the observed suppression of high $p_T$ particles, particularly
$\pi^0$'s from the PHENIX collaboration. X.N. Wang and others have
pointed out that high p$_T$ jets could be used as probes of the early
stages of heavy ion collisions\cite{xnwang}. High p$_T$ partons are
the result of hard collisions which happen very early in the
collision.  Hence as they traverse the fireball, their characteristics
can give us information about the system from the earliest times.  In
particular, the energy loss of these partons should be much greater in
a system of deconfined quarks and gluons, since they are directly
exposed to the color charges of the medium. Because the jets resulting
from these high p$_T$ partons are difficult to isolate from the soft
background, it was suggested that leading particles could be used
since they typically take about half the energy of the jet. Early
calculations simply assumed a constant energy loss per
unit length. More detailed QCD calculations\cite{dokshitzer} indicate
the energy loss actually depends on the total path length of the
parton through the fireball. This rather non-intuitive fact is a
result of interference effects between the soft gluons radiated from
the parton as it looses energy. The energy loss in a QGP can be
extremely high - greater than 10 GeV/fm at the highest energy
densities presumed to be created in these collisions.  However, since
the system expands and the lifetime of the high density stage is
rather short, the average energy loss of a parton is considerably
less.

Before looking at the data, it is worth briefly describing the
assumptions used to scale pp collisions to Au-Au collisions. There are
two basic scaling methods. The first, and most familiar to heavy ion
physicists is scaling with N$_{part}$, appropriate for are thermal
distributions coming from soft processes. Momentum distributions are
usually plotted as a function of the transverse mass, $m_T$, a
practice coming from that fact that all particles regardless of mass
could be described as a universal exponential in $m_T$ for lower energy pp
collisions. In heavy ion collisions exponentials tended to describe
the shape of the spectra, however the slopes diverge with the
heavier particles having larger inverse slopes. This was attributed to
the effect of hydrodynamic flow tending to give a higher transverse
momentum to heavier particles. These processes reflect the bulk nature
of heavy ion collisions.  Much of the data from the AGS and SPS scale as a
function of the number of participants, N$_{part}$, which in a central
Au-Au collision is 2*197. A notable exception to this was the
production of strange particles which scaled faster than this, and has
been attributed to the enhancement of strangeness in heavy ion
collisions. The second scaling method is more appropriate for hard
processes - i.e. jet production, and is of course more familiar to
those working at high energies. Here spectra can be described as a
power law p$_T$. There are no collective effects since one
assumes that there is essentially one hard interaction. Particles
originate from jet fragmentation. Here the scaling to heavy ion cross
sections from pp goes as the number of binary collisions - i.e as the
number of hard collisions. A central Au-Au collision has about 1000
binary collisions. Analysis of multiplicity distributions
provide evidence for a part of the cross
section scaling as the number of binary collisions\cite{binary}.

\begin{figure}[htb]
  \begin{center}
   \mbox{\epsfxsize 5in \epsfbox{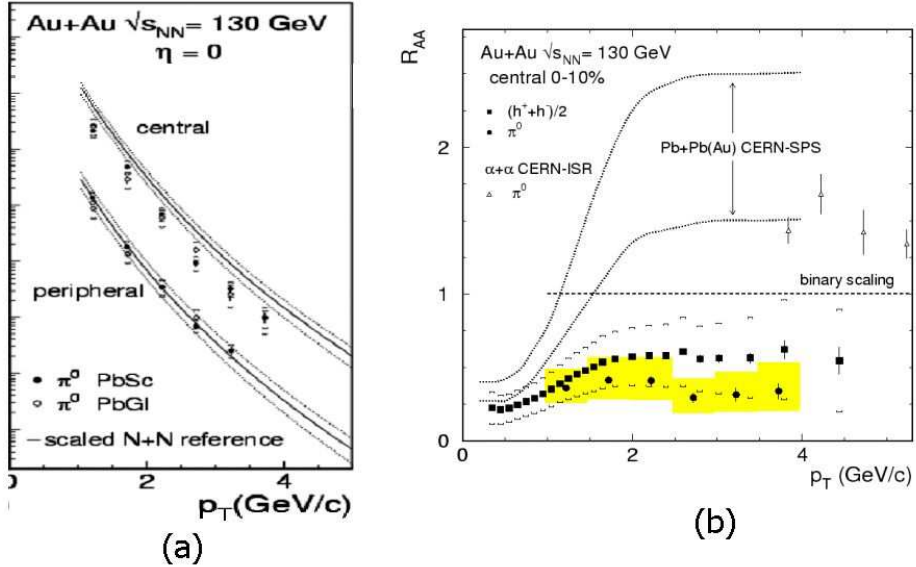}}
  \end{center}
\caption{(a) Transverse momentum distributions for $\pi^0$'s for 
 0-10\% 
central and 60-80\% peripheral events from the PHENIX collaboration as
compared to that expected from binary scaling from pp collisions.
(b) The ratio of the transverse momentum distributions divided
by the expectation from the binary scaling of pp collisions for both
charged and neutral hadrons. The errors bars indicate statistical
errors; systematic errors are indicated by the bands.}
  \label{fig6}
\end{figure}

Figure \ref{fig6} shows the $\pi^0$ p$_T$ distribution together with the
expected value from the scaling of the pp cross section with the
number of binary collisions\cite{phenix_prl} for both peripheral
collisions and central collisions. As one can see, the the scaling
works well for peripheral collisions. In the central collisions,
however, the data is over-predicted. Charged particles show a similar
behavior, with some complications, as I will describe in a moment.
The effect can be highlighted by dividing the data by the scaled pp
distribution as in the right panel, shown now both for $\pi^0$'s and
charged particles. In both cases the result is considerably below one.
Also shown is the range of results compiled from data at the SPS.  The
data from the SPS below $\sim 1.5 GeV$ also lies below one, since the
production mechanism for these lower energy particles scales as
N$_{part}$.  These are the vast majority of particles. We are concerned
with the small number of high p$_T$ particles above $\sim 2 GeV$. (The
exact value of ``high'' p$_T$ is open to question). At the SPS what we
see is an enhancement of high $p_T$ particles coming from quark
scattering - the well known ``Cronin'' effect. What is seen at RHIC is
quite the opposite.  There is a strong suppression above 1.5 GeV for
both the charged and neutral particles.  Such a suppression may be the
first indication of a large energy loss in a QGP.

\begin{figure}[htb]
  \begin{center}
   \mbox{\epsfxsize 6in \epsfbox{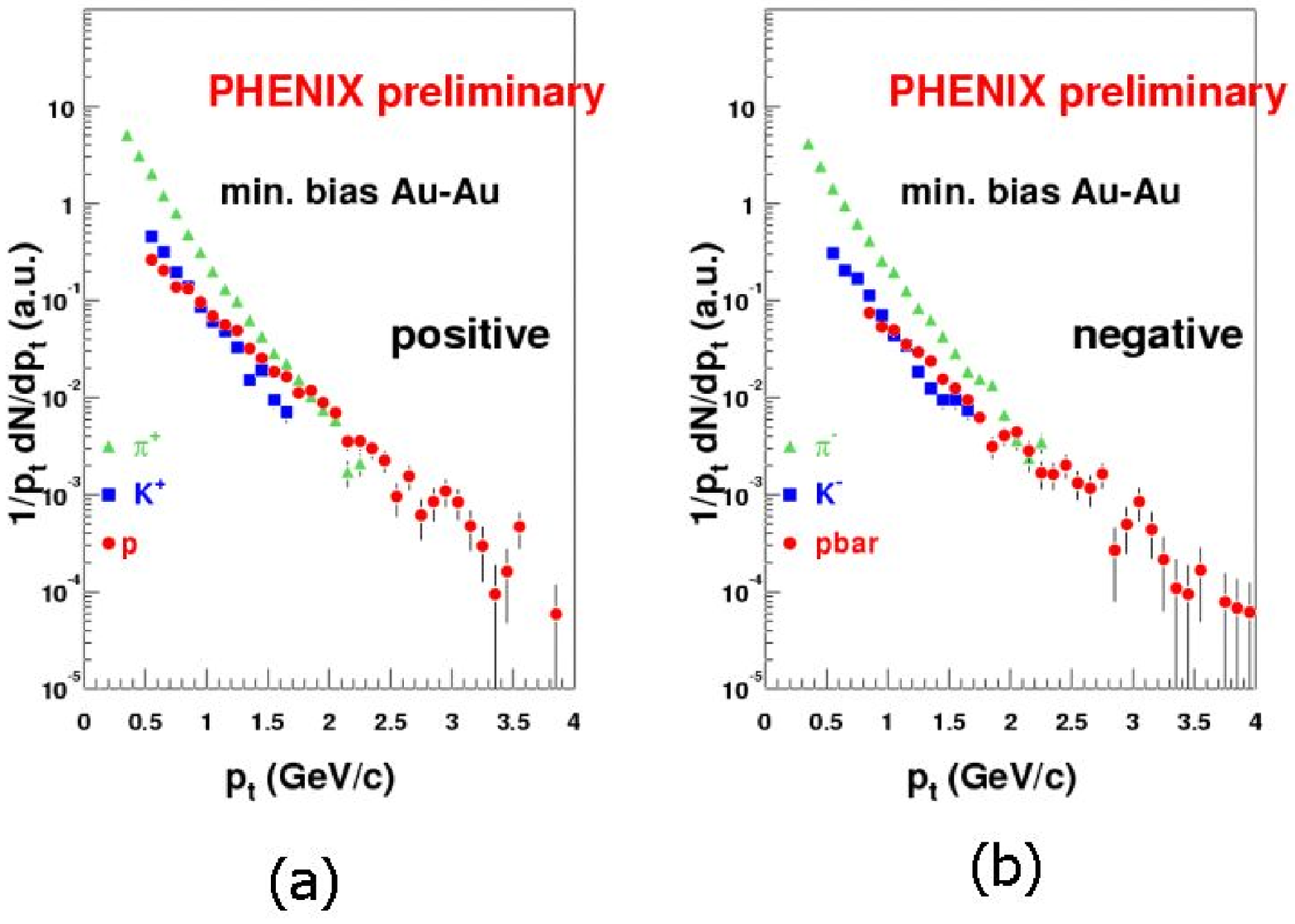}}
  \end{center}
\caption{Transverse momentum spectra for identified charged hadrons.}
  \label{fig7}
\end{figure}

One may then inquire as to the difference between the charged and
neutral spectra. Figure \ref{fig7} shows the charged spectra for
minimum bias events from PHENIX, now using particle ID to the extent
possible. One can immediately see the problem. At momenta above 2 GeV,
protons and anti-protons dominate the spectra! How can this happen?
This certainly is not characteristic of the distributions of leading
particles in jets, where only 10\% is a baryon\cite{SLD}. It may be that
collective effects such as radial flow are actually pushing the
baryons to a higher p$_T$ and we are comparing particles which come
from different sources - i.e. jet fragmentation after energy loss for
the pions, and low energy protons and anti-protons which have been
accelerated due to radial flow effects. At the same time the high
p$_T$ partons themselves loose energy and their daughter pions will
have a more thermal spectrum then would be characteristic of jet
fragmentation.  One of the unanticipated implications for the future,
is that particle identification at high $p_T$ will be critical to the
upgrades considered at RHIC.

Data on leptons and photons will be forthcoming.  Already PHENIX has
published the first results of electron production giving limits on
open charm. This is of particular interest because of the enhancement
of intermediate mass dileptons at CERN. However, since charmed quarks
themselves pass through the medium just as any other quark, they
should loose energy as well and be suppressed; though this effect
could be minimized due to a ``dead cone'' effect\cite{deadcone}.
Within the next year we should begin to see results on J/$\psi$
production in both the electron and muon channels, low mass vector
mesons to the di-electron channel, and perhaps thermal dileptons and
direct photons.

\section{What Would We Like to Know and what Should We Measure?}\label{measure}
If the view espoused here of the high p$_T$ spectra prevails, it will
provide vindication of the model described previously and lead to more
conclusive results regarding the possible formation of a Quark Gluon
Plasma.  Perhaps, as I have indicated, it raises more questions than
it answers. As a part of the task of putting together the Long Range
Plan for Nuclear Physics several meetings were held to consider the
future of of the RHIC program. In doing so, the following four major
questions were formulated.

\begin{itemize}
\item {\it Evolution of the System:} In relativistic heavy-ion
  collisions, how do the created systems evolve? Does the matter
  approach thermal equilibrium? What are the initial temperatures
  achieved?
\item {\it Deconfinement:} Can signatures of the deconfinement phase
  transition be located as the hot matter produced in relativistic
  heavy-ion collisions cools?  What is the origin of confinement?
\item {\it Chiral Symmetry Restoration:} What are the properties of
  the QCD vacuum and what are its connections to the masses of the
  hadrons? What is the origin of chiral symmetry breaking?
\item {\it Matter at the Highest Energy Densities:} What are the
  properties of matter at the highest energy densities? Is the basic
  idea that this is best described using fundamental quarks and gluons
  correct?
\end{itemize}

In what remains of this paper, I would like to take each of these
questions in turn and make some suggestions about measurements
which might help us answer these questions, and point out what 
upgrades will be necessary. In any set of such suggestions, there is
a danger of compartmentalizing certain results to answering certain questions.
In reality our subject is rich with overlapping subjects - measurements meant
to answer questions about the later hadronic part of the collision may lead to
information about the early stages of the collision.

I will also give more detailed examples of what might be done in three
particular measurements - low mass vector mesons, onium suppression,
and jet suppression - keeping in mind the statements I have made about
precision measurements.  I will conclude with a suggested set of
priorities for our community.

\subsection{Evolution of the System}

This is perhaps the question for which the suite of detectors at RHIC
are most capable of answering, at least for the later hadronic stage
of the collision.  We already have a basic understanding of much about
the final freeze-out phase.  It appears that the final state hadrons
reflect a both chemical and kinetic thermal distribution. We have
first estimates of the kinetic and chemical freeze-out temperatures,
chemical potentials, and expansion velocities.  There is much left to
understand, however, we have the tools in hand to do much of the work.
Detectors are well equipped to study HBT, flow and particle yields and
spectra.

Thanks to improvements in theory, we may be close to some
understanding of the initial stage of the collision as well.  Detailed tests
of the CGC model will be made in future eA machine proposed to be done
at RHIC and in pA collisions with upgraded detectors for looking at
collisions involving small x partons.

What is missing is a clear understanding of the stages, 
after the formation of the CGC particularly related to the
thermalization of the initial gluon distribution. Does the strength of
the elliptic flow signal really imply that thermalization occurs as
early as 0.6 fm? If so, it may be that we already have the answer in
hand but it is important to gather more information.

One of the possibilities for examining this stage of the collision is
though real and virtual photons from quark-anti-quark annihilation,
materializing as electron or muon pairs radiating from the hot, dense
QCD matter. While such radiation is emitted at all times during the
collision, the reaction dynamics favors emission from the hottest part
of the colliding system.  Thus, measurement of the distribution of the
blackbody thermal radiation will give us information on the initial
temperature, and possibly the black body nature of the spectrum. If so,
these photons would play the same role for us that the Cosmic
Microwave Background Radiation did for our understanding of the early
universe.  The background to such a signal is formidable since photons
and electrons are copiously produced from other sources such as
$\pi^0$ decay.  Upgraded detectors designed to reject such backgrounds
will be necessary. Systematic analysis, and variation of the initial
conditions will be required to solidify the interpretation.

\subsection{Deconfinement}

As mentioned previously, the suppression of high p$_T$ hadrons from
the energy loss of partons as they pass through a deconfined medium is
an important signal of deconfinement.  This may be enough to establish
the effect, however it will be important to be more quantitative.
This can be done by making measurements of direct photons produced
opposite high transverse momentum hadrons.  Since the photon recoils
against the quark jet, and since it does not suffer energy loss in the
deconfined medium, the photon serves as a indicator of the initial
transverse momentum of the jet.  This will provide a means to make
careful, quantitative measurements of the energy loss.  One
interesting possibility is to flavor tag the high transverse momentum
hadron. A leading K$^-$ with no valence quarks is more likely to come
from a gluon jet. This would allow one to measure the difference in
the energy loss between gluon and quark jets. Gluon jets are expected
to loose energy at twice the rate of quark jets in a deconfined
medium. These measurements will require high luminosity because of the
low cross sections of such high p$_T$ events. In addition detectors
will be required to handle the large luminosities, as well as to
identify hadrons to high p$_T$.

J/$\psi$ suppression is another well-known signature of deconfinement.
PHENIX will be able to measure J/$\psi$ production in both the muon and
electron channels. STAR will have access to the electron channel
within the next several years as their electromagnetic calorimeter is
completed providing a second measurement of this signature. 
One of the critical measurements that must accompany the
measurement of the J/$\psi$ is that of open charm production
to see whether open charm is enhanced or comes from the ``melting'' of the
J/$\psi$. To do this,
specialized vertex detectors must be added with the position
resolution that would allow a measurement of the charm vertex
separated from the original event vertex. STAR, PHOBOS and PHENIX have
all embarked on R and D programs to construct such an upgrade.

\begin{figure}[htb]
  \begin{center}
   \mbox{\epsfxsize 4in \epsfbox{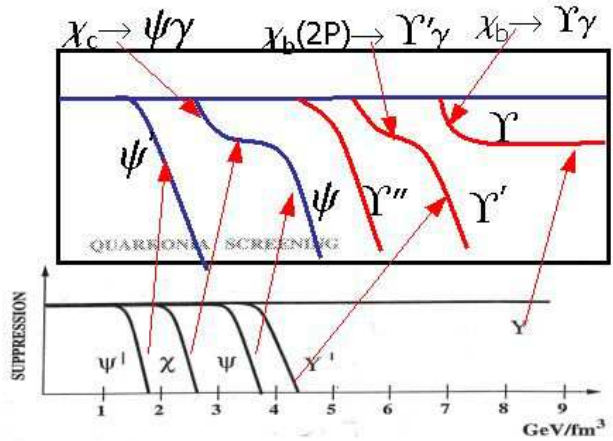}}
  \end{center}
\caption{Bottom Plot: Some possible ``melting'' points for the onium states. 
Top plot: the corresponding ratio to the unsuppressed values that one would 
observe experimentally.}
  \label{fig10}
\end{figure}

The J/$\psi$ is but one of the vector mesons in the charm family. The
excited states of the J/$\psi$ as well as the $\Upsilon$ family will
all exhibit some degree of suppression. The suppression of the
associated $c\overline{c}$ and $b\overline{b}$ states, 
can also be observed since they decay to
the detectable vector mesons.  Each of these states will "melt" at a
different temperature (figure \ref{fig10}). In fact the $\Upsilon$ will
be used as a control since it should not be suppressed at all at RHIC
energies. By varying the temperature of the system though changes in
beam energy and species, one can change the pattern of suppression of
the various states. Not only would this be a convincing signature of a
phase transition, it would give a good measure of the actual energy
density.  

It appears that the densities are high enough, even in mid-central
collisions where the reaction plane can be obtained, that there is a
QGP formed albeit with a smaller volume then in very central
collisions. This then raises the possibility that one can actually do
tomography on the collision region and map out its density profile to some
degree in the following manner. One would measure the angle between
the reaction plane and the probe of interest, for instance a high pt
particle opposite a photon and measure the energy loss as a function
of the reaction plane.  The path length perpendicular to the reaction
plane is longer than in the direction of the reaction plane because of
the almond shape of the collision region in semi-central collisions.
As an example of a particularly interesting question that this could
address is to see the relationship between energy loss and path length
on the one hand and energy density on the other by examining the
differences in energy loss as a function of the path length. This
would enable one to check the path length squared dependence of the
energy loss, mentioned before.

The measurements mentioned in this section would require measuring small
cross section processes in many bins of centrality, p$_T$ and reaction
plane. This demands an increase in the luminosity as well as 
large acceptance detectors capable of measuring leptons, photons, 
high pt particles with particle identification.

\subsection{Chiral Symmetry Restoration}
Chiral symmetry is broken through the creation of a vacuum scalar
condensate that couples to hadrons and provides most of their mass.
The challenge for RHIC experiments is to search for evidence of
in-medium mass changes of the low mass vector mesons associated with
the restoration of chiral symmetry. A direct measurement of the mass
of light vector mesons such as the $\rho$, $\omega$, and $\phi$ is
possible since they decay rather rapidly within the fireball created
at the time of collisions and before hadronization. The decay to
di-electrons is particularly interesting since electrons should not be
re-scattered in the medium and their invariant mass should reflect the
mass of the vector meson in the altered vacuum state.  Since some
fraction of the vector mesons decay outside the medium (in the case of
the $\omega$ some 70-80\%), these can be used as a
calibration point for the measurement.  The fraction exhibiting a
shifted mass should change as a function of the transverse momentum
and the size of the central fireball.  This would be a particularly
dramatic signature of the altered vacuum. As in the case of the
thermal di-electron signal, a major upgrade will be needed to reject
background for detection of the $\rho$, the shortest lived, and hence
the broadest of the vector mesons unless it is strongly enhanced which it
may well be.

\begin{figure}[htb]
  \begin{center}
   \mbox{\epsfxsize 6in \epsfbox{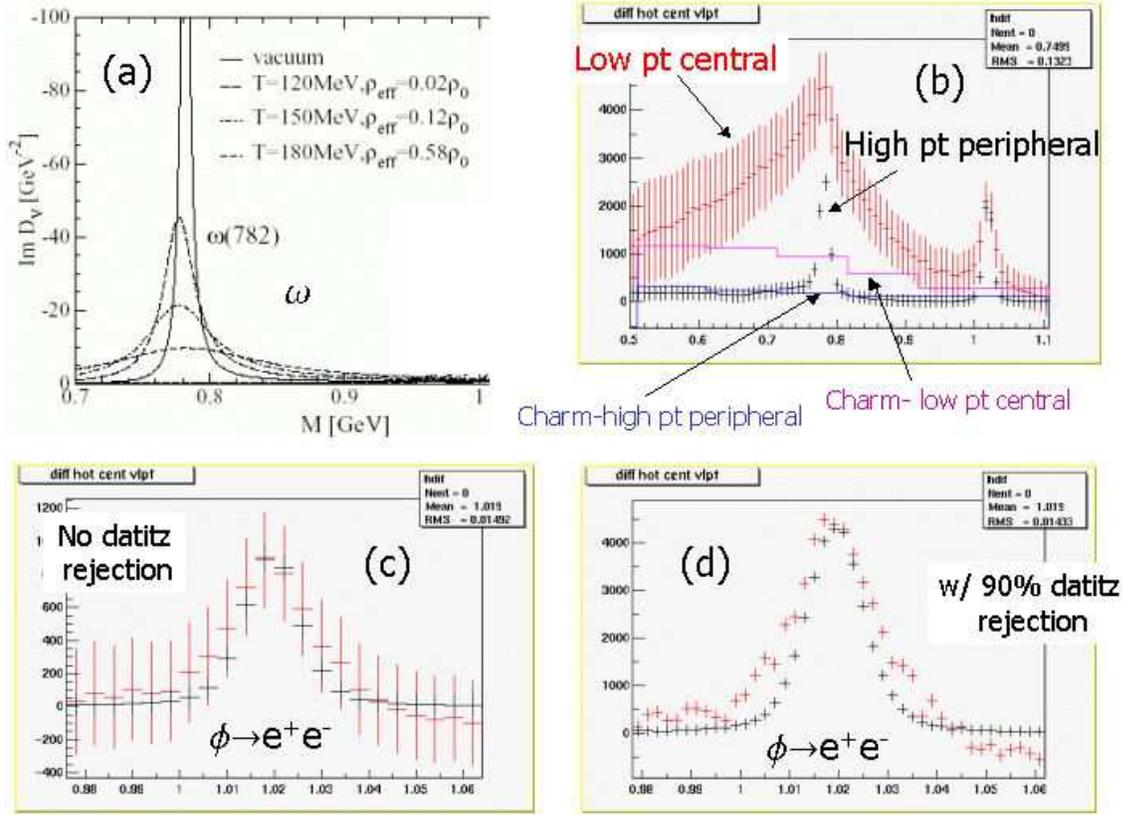}}
  \end{center}
\caption{(a) The broadening of the $\omega$ resonance for varying 
temperatures in the Rapp-Wambach model (b) Simulation of the di-electron 
signal in the $\rho-\omega$ mass region from high $p_T$ peripheral 
collisions and low $p_T$ central collisions without Dalitz rejection
as explained in the text assuming the Rapp-Wambach model (c) same as figure b 
for the $\phi$ as explained in the text without Dalitz rejection. Lighter lines indicate the signal in low $p_T$ central collisions, darker lines
indicate the signal in high $p_T$ peripheral collisions. (d) same as 
figure c assuming 90\% Dalitz rejection.}
  \label{fig8}
\end{figure}

The presence of a phase transition as the system cools is also
expected to cause fluctuations, which may survive the hadronic phase
as fluctuations in particle number and type. Fluctuations and droplet
formation are of particular interest, since similar processes may
account for much of the large scale structure of the universe and the
inhomogeneities observed in the cosmic microwave background.  A
variety of fluctuations have been proposed as a signature of a phase
transition. If the transition is first order, the growth of hadronic
droplets and the shrinking of quark-gluon droplets may yield a lumpy
final state and large fluctuations in particle number . If the
transition is a smooth but sufficiently rapid crossover, domains of
misaligned chiral condensate may be formed.  If the transition occurs
near a critical point separating first order behavior from crossover
behavior, long wavelength fluctuations imprint unique signatures on
the momenta of soft pions. Experiments will search for such phenomena,
and correlate their appearance with other quark-gluon plasma
signatures.

The models of chiral symmetry restoration have followed essentially
two approaches. The first, typified by Rapp and Wambach's
model\cite{rappwam} assumes hadronic degrees of the freedom as the
staring point. The other, typified by Brown-Rho
scaling\cite{brownrho}, uses primarily quark degrees of freedom.  Under
the hadronic description, the restoration of chiral symmetry 
 means that the vector (e.g. the $\rho$) and
the axial vector (e.g. the $A_1$) will become degenerate when chiral
symmetry is restored. Typically the particle widths increase and the
mass does not go to zero. Brown and Rho however, assumed that all particle
masses scale with the chiral condensate which of course goes to zero
when chiral symmetry is restored. 
These two views are of course, seemingly contradictory - a
situation that is somewhat endemic to this subject.  The two
groups have made the conjecture that there is
a quark-hadron duality and that the two approaches will give the same
results if one did not have to resort to perturbation theory, but
somehow could calculate the results directly which, of course
is probably not possible, at least not analytically. The only difference
between the two approaches, then is simply the difference between
the sets of basis states or ``degrees of freedom'' that they choose
to use to solve the problem. 

Let us suppose that the chiral and deconfinement
transition occur at the same place. Let us also assume that the Rapp
Wambach model was correct below the phase transition, where at least
at very low temperature their set of basis states seem reasonable,
and Brown Rho scaling is appropriate above the transition in the sense
the the quark masses scale with the chiral condensate. The Rapp
Wambach Model predicts that as we approach the phase transition from
below the vector and axial vector particles widen and become
degenerate. Above the phase transition, the quarks deconfine and
presumably the light quark masses will go to their bare quark masses -
essentially zero on the scales which we are considering. At the phase
transition, somehow these pictures must merge. It may be that neither
set of basis states are the correct ones to use near the phase transition.
If one was able to find the correct basis states, one
may actually be able to do the calculation in the region of interest.

A similar situation was faced by condensed matter theorists in trying
to explain super-conductivity. Bardeen, Cooper and Schrieffer managed
to guess the correct set of basis states to use - Cooper pairs.  
Presumably they could have constructed their theory using electrons as
the basis states. The theory would have been just as good but it would
have been hopelessly complicated.  Finding the right set of basis
states is one of the major steps to solving such a problem. In the
case of QCD, it may be that neither the hadronic nor the quark basis
are the best set of basis states to use near the phase transition. It
also may not be possible to guess the right set of such states with
knowledge available to theorists now. What is needed are experimental
results, which then can used as a foundation for making a reasonably
educated guess.  Of course we can approach this problem in a brute
force manner using lattice gauge theory on massive computers, however
finding the right degrees of freedom or basis states gives one a much
better qualitative understanding of the phenomena.  This is one of the
instances in which the interplay between theory and experiment is the
most productive.

The detection of low mass vector mesons, particularly the $\rho$,
decaying to di-electrons is a challenging task for similar reasons to
the the problem of identifying the thermal di-electron signal
mentioned previously- the backgrounds are fierce from $\pi^0$ Dalitz
decays and photon conversions.  A scheme of rejection of
Dalitz and conversion pairs must be implemented.

The $\rho$, in addition to being useful for the understanding of
chiral symmetry restoration, can also be used as a clock of the mixed
and hadronic phases of the collision. Since the lifetime is so short,
the $\rho$ decays and is regenerated several times during the
collision - for instance from $\pi \pi$ interactions.  Since electrons
do not interact strongly they can emerge from the collision relatively
unhindered. A long mixed and hadronic phase should enhance the $\rho$
to di-electron signal in a matter which yields the lifetime.

In order to study some of these effects, and the need for Dalitz
rejection, a simulation was done using as input, the model of Rapp and
Wambach which includes the restoration of chiral symmetry and the
enhancement of the $\rho$ due to the aforementioned effect. Figure
\ref{fig8}a shows the resonance shape of the $\omega$ meson for a
variety of temperatures.  One can see the slight mass shift, and the
broadening of the line shape.  A run of $10^9$ central events (about 2
years of running) was assumed.  Figure \ref{fig8}b shows the signal
for the $\rho$, $\omega$ and $\phi$ assuming the current PHENIX detector
which has no Dalitz rejection. The first thing to note is that if the
enhancement of the $\rho$ is as strong as the Rapp-Wambach model
indicates, the signal will be clearly measurable even with the current
configuration of the PHENIX detector without Dalitz rejection. 
To illustrate the difficulties, however, we turn to the
$\phi$ which shifts in a manner similar to the $\omega$. Figure
\ref{fig8}c shows a comparison of the signal from low-pt central
events where the effect should be the strongest, to the high-pt
peripheral events where it should be essentially normal.  The
statistical error bars are very poor due to the large subtraction
which must be performed due to the background.  Clearly, one would not
be able to deduce any difference. A 90\% Dalitz rejection was assumed
for figure \ref{fig8}d, and a clear effect of width broadening due to
chiral symmetry restoration can be seen.

There are essentially two schemes being considered for Dalitz
rejection. One of the methods contemplated is that of a hadron blind
detector with large acceptance which has a high probability of
detecting both particles of a Dalitz pair. This typically involves a
field free region near the vertex so that the low momentum partner is
not bent out of the acceptance of the detector. Since the detector is
sensitive to electrons which have a relatively low multiplicity, electrons
and positrons can be reconstructed  and rejected if they are
consistent with a Dalitz of conversion pair. Such a scheme could even
be implemented on-line. The second involves the use of a vertex detector
together with the ability to do good particle identification. Such a
detector could measure low momentum electrons since it is so close to
the vertex. The background pair could then be rejected in a similar
manner to the hadron blind scheme.

\subsection{Matter at the Highest Energy Densities}

There is a great deal to learn about hadronic interactions at very
high energies where we are probing the very low x region of the hadron
wave-function. The CGC is critical to understanding these process and
an understanding of the gluon densities at these high energies forms
the basis for the calculations of the initial conditions in
nucleus-nucleus collisions. Critical experiments to perform will
include pA and eA collisions.  A future project at the RHIC Collider
will be the addition of an electron ring to study eA collisions.
In the near future we can begin to use pA collisions. Proton-nucleus
collisions are necessary component of the heavy ion program as well,
since pA as well as pp collisions serves as a basis for the
normalization of AA collisions to provide an understanding of
phenomena such as nuclear shadowing.  An increase in luminosity is
necessary for this task since the same rare signals must be detected.

Signals which can be used to measure the relevant gluon distributions
primarily come from gluon fusion processes. These include the
production of heavy quarks and W bosons and direct photons. Anti-quark
distributions can be probed via Drell Yan. Saturation will affect the
rate and the transverse momenta of these probes. Processes which
involve very low -x partons will occur in the very forward portion of
the detector and upgrades to lepton, photon and micro-vertex detectors
which can cover pseudorapidities as high as $\eta \sim 5$ are required.

\subsection{What is Needed?}\label{needed}

We are now left with the task of setting the priorities for the future
of the RHIC program. One of the important aspects providing enormous
strength to the program is a redundancy in the capabilities of the
detectors. The different detection and analysis techniques as well as
different regions of acceptance can provide a cross check and
corroboration of critical signatures lending credibility to the
discovery of new phenomena. In particular, STAR is now adding 
electromagnetic calorimetry over a large acceptance,
providing balance to PHENIX's aggressive electron and muon
identification over a somewhat smaller region.

R and D for the detector upgrades has begun.  Funding must be provided
for this. Upgrades which will be added should and does include a
program for Dalitz rejection, high p$_T$ particle identification,
and micro-vertex detection for heavy quark production. 
Weaknesses of the various detectors are being addressed - for
instance in the addition of electromagnetic calorimetry for STAR as
just mentioned, and the addition of 4$\pi$ tracking for the PHENIX
detector to enhance jet detection capabilities.  STAR already has a
good acceptance at very forward rapidities. PHENIX is contemplating
the addition of very forward detection of muons and photons as well as
a micro-vertex detector in the forward region, though these projects
are not as well formulated as some of the others.All detectors must of
course enhance their data-acquisition capabilities to enable them to
take advantage of the planned luminosity upgrade.

\section{Conclusions}\label{concl}

RHIC provides us with a unique capability to understand QCD
interactions, both at high temperatures and in bulk, which was the
primary focus of this paper. It will also study pp, pA, and eA 
collisions in which we can study the quark and gluon distributions
themselves as well as the spin of the nucleon in the very rich
polarized proton program which I have not mentioned at all.

I remain optimistic about many of the ideas as explained in this
paper, particularly the Colored Glass Condensate.  However it must be
emphasized that it is far too early to assume that it is a well
established theory. Only further experimental evidence can tell us
whether it is right or wrong. In fact it may well be that the CGC
model turns out to be incorrect, however, the important point 
is that the future of the RHIC experimental program (and
LHC) is very bright.  The future of theory in the field is progressing
rapidly as well and hand in hand, theory and experiment will work
together such that perhaps in the next 5-10 years, our picture of
high-temperature QCD will be much more complete giving us an
understanding of the nature of the phase transitions and the nature of
confinement and hadronic mass-which I remind the reader makes us most
of the ordinary matter we see around us.

\section*{Acknowledgments}
I would like to thank the numerous theorists who have patiently
explained much of this material to me and corrected many of my
misconceptions. I would like to particularly thank Larry Mclerran who
read much of the introductory material and reigned is some of my
enthusiasm for the absolute reliability of QCD calculations for the
energies available at RHIC and reminded me that our task was not so
much to create or ``discover'' the Quark-Gluon-Plasma but to
understand it. Any errors remaining in the text are mine.

\end{document}